# Advanced Physical-Layer Technologies in VHF Data Link Communications

Hosseinali Jamal, *Member IEEE*, David W. Matolak, *Senior Member IEEE*

University of South Carolina
Department of Electrical Engineering
Columbia, SC
hjamal@email.sc.edu, matolak@cec.sc.edu

*Abstract*— The VHF aviation band is preferred for narrowband and long-distance communications due to its modest channel attenuation. Hence, this band has been widely used for aviation voice communication using analog communication systems for decades. However, due to the rapidly increasing number of flights and high usage of VHF channels, the VHF spectrum is becoming much more crowded, and use of analog waveforms will likely not maintain the required performance and quality of service in the future. Therefore, digital communication systems have been considered and studied due to their larger spectral efficiency. Notably, digital communication systems that have been proposed for VHF are broadband VHF (B-VHF) using multi carrier-code division multiple access (MC-CDMA), and VHF data link modes 2 and 3 (VDL2/3) using differential 8-state phase shift keying (D8PSK) modulation. Compared to B-VHF, VDL2/3 has received more attention due to its simplicity and more constant amplitude waveform, yielding lower peak-to-average power ratio (PAPR) and hence better energy efficiency. Recently, advanced VHF digital link (A-VDL) was proposed for VHF [1]. This scheme enables use of essentially the same platform as VDL except for the physical layer processing, including modulation. The proposed A-VDL, following digital video broadcasting satellite second-generation (DVB-S2) standard, uses amplitude and phase-shift keying (APSK) modulation with higher modulation order than VDL, hence providing higher spectral efficiency than VDL. Compared to the widely used quadrature amplitude modulation (QAM), APSK is more resistant to amplifier amplitude and phase distortions. Thus, APSK has become of interest for satellite communications, as well as VHF communications in A-VDL. In this paper, we investigate other advanced technologies such as channel encoding techniques used in DVB-S2, low-density parity-check (LDPC) codes, more efficient standardized voice encoders, as well as better pulse shaping filters than the classical square-root raised-cosine (SRRC) filter used in VDL and A-VDL. Via simulations and analysis, we compare the proposed scheme's link margin, PAPR, and spectral efficiency compared to VDL and A-VDL, which both use Reed Solomon (RS) encoding. In addition, as another way of generating the same VDL waveforms (or possibly other single-carrier aeronautical band waveforms), we investigate the single-carrier type waveform used in cellular LTE and 5G uplink communication links: the discrete Fourier transform-spread OFDM (DFT-s-OFDM), and discuss how we can take advantage of using the same LTE and 5G hardware resources.

*Keywords— A-VDL, DVB-S2X, LDPC, APSK, DFT-s-OFDM*

## I. Introduction

Two fundamental modes of VHF communications currently exist: voice and data. Analog voice has been used for many years within the VHF aeronautical communications frequency band 117.975-137 MHz. The analog voice on-the-air is based on conventional double-sideband amplitude modulation (DSB-AM) with no carrier suppression. In comparison to the digital communications options, the analog VHF channelization limits the channel capacity significantly. It is recognized that by only using the analog communications, there will be a shortage of assignable aeronautical VHF communications channels in some regions of the world [2]. The shortage of VHF channels and resources could seriously affect the air traffic services (ATS) communications and aeronautical operational communications (AOC) needed to cope with current and future air traffic communication needs. Thus, some researchers have recently re-opened investigations into use of new techniques in the VHF band [1].

Digital voice has been proposed as a solution to increase the spectral efficiency but hasn't been placed into service until recently. Currently used systems such as analog data aircraft communications addressing and reporting system (ACARS) is applicable, using an audio minimum-shift keying (MSK) modem: ACARS over aviation VHF link control (AVLC) or AOA. The ACARS avionics sends digital voice communication with data rate of 2400 bps. Despite the use of spectrally efficient MSK, full advantage of digital communication is not realized because of the necessity of reducing capability to enable interfacing with existing DSB-AM transmitters and receivers [2]. Since ACARS is meant to be phased out, VHF digital link (VDL modes 2 and 3) will probably replace it, but currently both AOA and "Plain Old ACARS" or POA is used in many commercial aircraft.

Air-ground propagation characteristics of the VHF band generally allow transmission and reception in line of sight (LOS) conditions, with maximum LOS ACARS range for an enroute aircraft at an altitude of 30000 feet about 250 nautical miles (nmi); of course, the range decreases at lower altitudes. The maximum range for VDL is 200 nmi [2]. VDL mode 2 (VDL2) and 3 (VDL3) are supported with 25 kHz channels. There is also VDL2 channel specification with 8.33 kHz bandwidth in some regions. It is required that the VDL2 transceivers must be capable of tuning to any of the 760 channels of 25 kHz in the band 117.975-137 MHz. Details about the VHF channelization can be found in [2] (appendix M). VDL 2 has been mainly proposed and designed to support some ATS such as controller pilot data link communications (CPDLC) supported by line of sight (LoS) communications.

Data in VDL typically consists of messages to/from pilots from/to ATS or AOC. Examples of messages are route clearances to/from air traffic control (ATC), and information between the aircraft and airline. VDL2 is appropriate for aperiodic traffic, which means the entire message is ready before transmission of individual message packets begins, hence VDL2 is not prepared for real-time applications such as real-time digital voice. In contrast, VDL3 provides datalink and real-time voice operation.

Despite using some advances of digital communications, VDL is more-than-a-decade-old technology and does not employ full advantages of current techniques used in modern communication systems and standards such as DVB-S2X, cellular 4G, and 5G. In [1], the authors investigated several advance-VDL (A-VDL) schemes and studied coherent modulations such as APSK (used in DVB-S2X), with lower coding rates than used in VDL[1]. According to their analysis, for the same symbol rate as VDL (10.5 ksps), using 16-PSK and lower code rate of $r = 3/4$ we can improve the link margin (LM) relative to VDL by approximately 8 dB and achieve the same information bit rate (after decoding). The use of the same symbol rate translates to the same RF channel bandwidth. Using even higher order modulation, e.g., 256-APSK with higher symbol rate (20 ksps) we can improve the spectral efficiency of VDL by order of 3 but losing approximately 7 dB in LM. The authors of [1] do not explain their LM analysis for comparing with VDL. In LM analysis we should compare the standard required energy per bit at the required bit error rate (BER) for the defined coding scheme, considering the peak-to-average power ratio (PAPR) and any average power constraints of the transmitted waveform.

In this paper, we build on the work of [1]. Other than APSK modulation, we investigate other advanced schemes such as the channel encoding used in DVB-S2X and 5G: low-density parity-check (LDPC). We also consider better pulse shaping filters than the classical square-root raised-cosine (SRRC) filter used in VDL and A-VDL, quadrature amplitude modulation (QAM) modulation. Some of these filters are used in 5G and new Wi-Fi standard releases. We also consider more standardized voice encoders (with better bit rate efficiency) which could increase the voice channel capacity. Via MATLAB Monte Carlo simulations we show the bit error ratio (BER) versus bit energy per noise spectral density ($E_b/N_0$) for different APSK and QAM modulation orders and compare these two modulations side-by-side. In our LM analysis, we also consider the PAPR of each of these modulations after showing the PAPR results.

The rest of this paper is organized as follows: in Section II, we provide a brief overview of VDL physical layer (PL), and describe the blocks used in VDL which we also consider in our simulations. In Section III we describe the advanced techniques used in our proposed advanced VDL schemes and explain the necessary changes required to VDL. In Section IV we provide simulation results and LM analysis comparing to VDL. Section V presents the discrete Fourier transform spread OFDM (DFT-s-OFDM) based implementation of the same VDL waveforms, which allows multiple access among users by assigning different non-overlapping Fourier coefficients (sub-carriers) to different users using similar 5G hardware resources.

## II. VDL PHYSICAL LAYER OVERVIEW

VDL transmission is single carrier half-duplex, time-division duplex (TDD) (uplink and downlink usually at same frequency), and similar to modern Wi-Fi, employs carrier sense multiple access (CSMA). The CSMA algorithm used in VDL2 allows the receiver to determine if the channel is idle by using an energy sensing algorithm. However, because the local noise floor is not a constant, an adaptive estimator is needed.

The VDL schemes use differentially encoded 8 phase shift keying (D8PSK) modulation, using SRRC filtering with roll-off factor $\alpha = 0.6$. The transmitted data is differentially encoded with 3 bits per symbol transmitted as changes in phase. Therefore, the resulting waveform has eight equal amplitude states with an angular spacing of $\pi/4$ radians (Figure 1). For the VDL2 channel bandwidth of 25 kHz, based on the spectral mask requirement [2], the data rate shall be limited to a nominal bit rate of 31500 bps, which corresponds to 10500 symbols/s. The forward error correction (FEC) for information channel coding is based on a (255, 249) Reed-Solomon (RS) encoding technique. More technical details regarding generating the VDL signal can be found in [2].

The VDL2 radio frame format is shown in Figure 2. The frame consists of two parts: a header training sequence and aviation VHF link control (AVLC). The AVLC data is protected by interleaving and a RS FEC, and a special FEC protects the header. The transmitter ramp-up consists of five symbols of 000. The purpose of this first segment of the training sequence is to provide for transmitter power stabilization and receiver automatic gain control (AGC) settling, and it shall immediately precede the first symbol of the unique word used for synchronization.

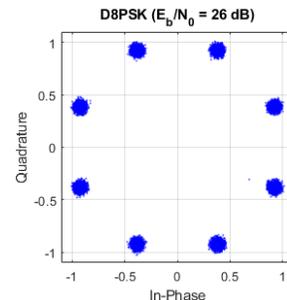

Figure 1. VDL2 D8PSK constellation.

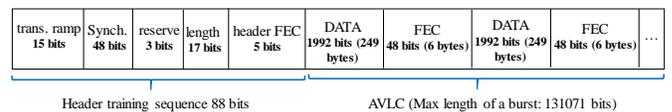

Figure 2. VDL2 frame format.

---
[1] The FEC scheme used in A-VDL is not mentioned in [1]. Assuming using the same LDPC coding scheme as we used in this paper, our link margins and those in [1] are similar for all modulations.

The synchronization pattern consists of the unique word: 000 010 011 110 000 001 101 110 001 100 011 111 101 111 100 010. The synchronization pattern is there to provide a known bit sequence at the beginning of the data burst to enable the receiver to do bit time synchronization and channel estimation. The reserved symbol consists of three zeros reserved for future definition. The 17 bits transmission length in the training sequence specifies the total number of bits that follow the header FEC, not including RS FEC and padded bits in the AVLC frame.

This 17-bit sequence is transmitted LSB first. Since the transmission length consists of 17 bits, the maximum length of any transmission is 131071 bits, not including RS FEC. The header FEC is a (25, 20) convolutional block code computed over reserved symbol and transmission length.

As mentioned in [2] the International Civil Aviation Organization (ICAO) standards and recommended practices (SARPs) require an uncoded BER of $10^{-3}$ and a coded BER of $10^{-4}$ for VDL2 operations. The common standard performance of $10^{-3}$ (uncorrected) BER has been established to simplify the design and qualification of ground and aircraft hardware supporting both VDL Modes. Based on VDL requirements the system should satisfy minimum of 5 dB LM to ensure link availability is no worse than the present DSB-AM analog voice system [2]. Note that following the VDL2 requirements, we will target the BER of $10^{-4}$ after decoding in our LM analysis as well.

### III. VDL Physical Layer Using Advanced Techniques

The A-VDL proposed schemes in [1] use essentially the same platform as VDL except for the physical layer processing, with the same channel bandwidth, with new modulation (and possibly coding scheme). The proposed A-VDL, similar to the modern satellite links DVB-S2X, uses APSK modulation with higher modulation order than VDL and higher symbol rates of 20 ksps (over same 25 kHz bandwidth but with smaller SRRC α), hence providing up to three times higher information bit rate than VDL (90 kbps bit rate). As described in [1], even higher data rates are possible for shorter communications ranges (i.e., less than 100 NM) for wider channel bandwidths (i.e., larger than 50 kHz).

APSK has been used in satellite communications, as satellite power amplifiers are often used at or beyond their compression levels to maximize their conversion efficiency and get as much output power as possible given the power-limited link. Compared to the widely used QAM, APSK is more resistant to the amplifier amplitude and phase distortions. Thus, APSK has become of interest for satellite communications, as well as VHF communications in A-VDL compared to QAM modulation, enabling use of the high power amplifier (HPA) closer to its saturation point.

Distortion from the nonlinearity of amplifiers beyond their compression levels can move the position of symbols around the constellation and cause errors in symbol detection and cause interference to other symbols. Looking at the constellation (see Fig. 5), the inner symbols (closer to the origin) are lower power symbols, and may not be distorted, while the outer points may drive the amplifier into compression and experience distortion. In APSK, the symbols are configured in concentric rings of constant amplitude (Figure 5). Unlike QAM, the states are configured in rings, with the intent that symbol points in each ring will react the same way to compression [3], [4]. This has two positive effects.

The first is that compression of the signal tends to have less of an effect on the Euclidean distance between states, and so the states are easier to distinguish from each other during demodulation. The second advantage of APSK is that it lends itself to pre-distortion. By varying the space between rings before transmission, it is possible to pre-distort the signal in a way that counteracts the effects of transmission distortion and thereby gets a better output. Hence, using dynamic pre-distortion (as in DVB-S2X), the signal received is monitored and measured, and the results are fed back to the pre-distortion circuitry for adjustment.

Therefore, in VDL links, if the requirements can allow use of the higher power levels of HPA along with pre-distortion techniques, APSK is a near ideal modulation which can improve the LM of the system as well as increase the spectral efficiency. In other words, APSK can be a valid alternative to other higher order modulations in all cases in which the nonlinear effects due to HPAs cannot be neglected.

In this paper, in our BER/PAPR simulations and LM analysis, we assume the HPA works in its linear region, and we compare the QAM with APSK with different modulation orders. For this assumption in LM analysis we compare the PAPR of each waveform generated with different modulation orders and apply a power back-off equal to $PAPR_0$ for an arbitrary yet fair assumption of $Pr(PAPR > PAPR_0) < 10^{-3}$. For coding in our proposed system, we consider the same LDPC used in DVB-S2X, with coding rate of 3/4. Note that this coding can provide any code rate among the following set, 1/4, 1/3, 1/2, 3/5, 2/3, 3/4, 4/5, 5/6, 8/9 and 9/10, but in our initial simulations and analysis we consider only one rate. The code block length of this LDPC code has length of 64800 bits. Therefore, considering this coding scheme, we can update the VDL frame structure as shown in Figure 3.

### IV. Simulation Results

In our simulations, referring to Figure 3, we only consider the AVLC part of communications, assuming perfect synchronizations, and an AWGN channel. We use two types of modulation, QAM and APSK with different orders of 16, 32, 64, and 256. For simulating the proposed VDL link, we follow the block diagram shown in Figure 4.

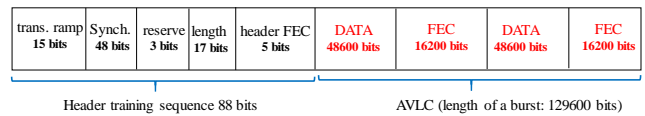

Figure 3. Updated VDL2 frame format using LDPC code.

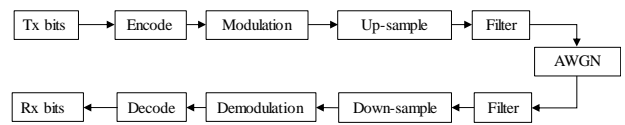

Figure 4. Communication system block diagram used in simulations.

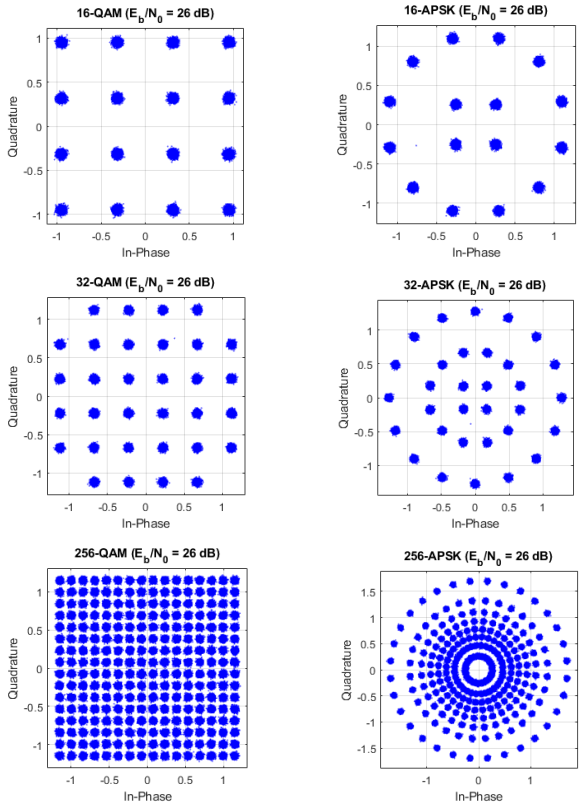

Figure 5. Constellations of QAM and APSK (DVB-S2X) using different modulation orders.

For the filter in these simulations we use our proposed filter described subsequently in the section on PSD results. We chose an up-sampling factor $N_{up} = 4$, and filter order $N_{up} K = 32$, where $K = 8$ is the length of filter in terms of number of symbols.

In order to check the constellations of the transmitted symbols before the AWGN block, we show the scatterplot of the waveform samples in Figure 5 for QAM and APSK modulations assuming $E_b/N_0$ of 26 dB.

Next we compare the BER performance of all modulation orders for both QAM and APSK, shown in Figure 6 and Figure 7, respectively. In these simulations, as another reference we also include the results considering convolutional encoding with the same coding rate of 3/4, and block length of 64800, using soft decision bits for decoding. Despite having worse performance than LDPC coding, one advantage of convolutional coding is the flexibility in choosing the length of the block codes which can be much smaller than LDPC code. The advantage of having smaller code words is the lower complexity and simpler frame format with adjustable packet lengths. Comparing Figure 6 and Figure 7, we realize the similar performance of QAM and APSK in smaller modulation orders of 16 and 32, but for larger modulation orders of 64 and 256 we notice that APSK has slightly better performance (approximately 0.2-0.3 dB) in this AWGN channel. In these figures we also include the performance of the reference D8PSK modulation used in VDL2, and a coherent 8-PSK option providing larger link margin due to lower coding rate of $r = 3/4$ and coherent demodulation.

In Figure 8 we compare the PAPR of the waveforms generated by different QAM and APSK modulation orders. We also include the PAPR of D8PSK (and 8-PSK) for link margin analysis of the systems. Looking at these PAPR curves, e.g., at probability of $10^{-3}$, we notice that the variation of the PAPR for all QAM modulation orders is less than that for APSK, with maximum variation of approximately 0.5 dB for QAM modulations and 2.4 dB for APSK. Another observation is that APSK PAPRs increase with modulation order, while for QAM, 16-QAM is similar to 32-QAM, and 64-QAM is similar 256-QAM.

For LM analysis comparing to VDL2, we consider the BER gain of each modulation order at the RTCA coded BER recommendation of $10^{-4}$, as well as the back-off power required compared to D8PSK for all QAM and APSK modulation orders. Therefore, we can calculate the LM gain relative to VDL2/3 as, LM = $E_b/N_0$ gain (at BER = $10^{-4}$) – required back-off (at $Pr(PAPR > PAPR_0) = 10^{-3}$).

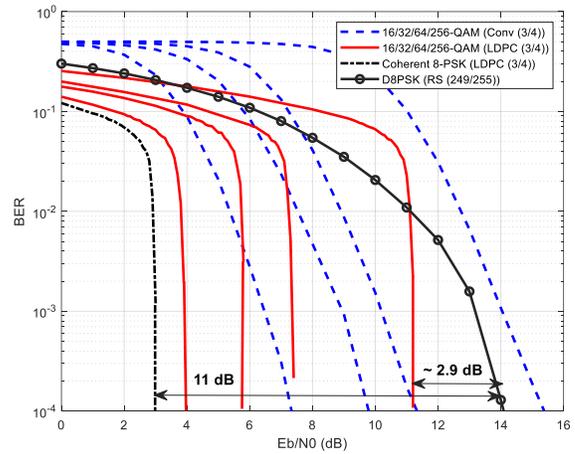

Figure 6. BER vs. $E_b/N_0$ for QAM. (Lower modulation orders to left, higher modulation orders to right.)

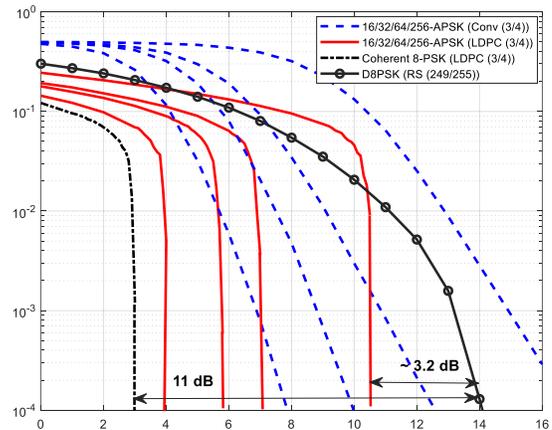

Figure 7. BER vs. $E_b/N_0$ for APSK. (Lower modulation orders to left, higher modulation orders to right.)

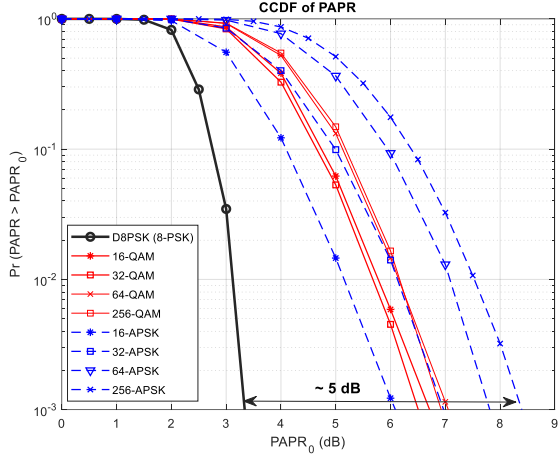

Figure 8. QAM and APSK PAPR comparison with D8PSK.

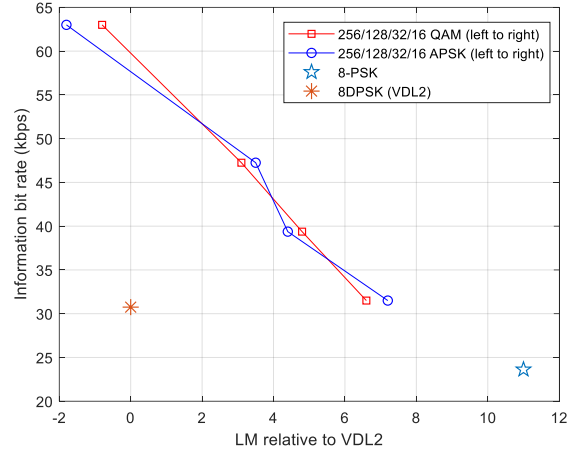

Figure 9. Information bit rate (after decoding) vs LM for different modulations.

Here the gain is over the D8PSK VDL scheme. In Figure 6 and Figure 7 as examples, we also show the $E_b/N_0$ gain for 8-PSK and 256-QAM/APSK modulations (with coding rate r = 3/4), which are 2.9, 3.2, and 11 dB for 256-QAM, 256-APSK, and 8-PSK, respectively. In Figure 8 we also show the required back-off power for 256-APSK, which is approximately 5 dB. In Table 1 we collect all these numbers for different modulations. We also provide the information bit rate (after decoding) assuming symbol rate of 10.5 ksps for each modulation.

To better compare these numbers, we plot the information bit rates versus LM for different modulations orders in Figure 9. According to these results, QAM and APSK almost perform the same, with 256-QAM outperforming 256-APSK by about 1 dB. These results show that with high modulation orders of 256 it is possible to increase the useful throughput by more than a factor of two, with only 0.8, 1.8 dB decrease on LM using QAM and APSK, respectively.

Table 1. Link margin and information bit rate analysis for different modulations (symbol rate 10.5 ksps).

| Modulation | $E_b/N_0$ gain (dB) | Required back-off (dB) | LM (dB) | Information bit rate (kbps) |
|---|---|---|---|---|
| D8PSK | | | | 30.75 |
| Coherent 8-PSK | 11 | 0 | 11 | 23.625 |
| 16-QAM | 10 | 3.4 | 6.6 | 31.5 |
| 16-APSK | 10 | 2.8 | 7.2 | 31.5 |
| 32-QAM | 8.1 | 3.3 | 4.8 | 39.375 |
| 32-APSK | 8.1 | 3.7 | 4.4 | 39.375 |
| 64-QAM | 6.8 | 3.7 | 3.1 | 47.25 |
| 64-APSK | 7 | 4.5 | 3.5 | 47.25 |
| 256-QAM | 2.9 | 3.7 | -0.8 | 63 |
| 256-APSK | 3.2 | 5 | -1.8 | 63 |

Using modulation order of 16 we approximately achieve the same throughput, with a gain of 7 dB in link margin. Further, using coherent 8-PSK modulation can provide 11 dB gain on LM with decreased throughput of approximately 23 percent due to the lower coding rate ($r$ = 3/4) than VDL2/3. Note that these results are for symbol rate of 10.5 ksps, and hence the same approximate channel bandwidth as VDL2/3. According to the analysis in [1], using higher symbol rates one can further improve the throughput while achieving smaller gains on LM compared to VDL2 (assuming same total transmit power).

In order to improve the spectral density of the transmitted signal within VDL2/3 channels we propose a different filter than the SRRC filter. The filter that we suggest is a design generated based on Parks-McClellan (PM) method [5]. For this filter we found that, the only passband and stopband cutoff frequencies that can achieve the Nyquist property (zero inter-symbol interference) are $f_{pass} = R/N_{up}$ and $f_{stop} = 3.4 f_{pass}$, where $R$ is the symbol rate and $N_{up}$ the up-sampling factor that we chose as 4 in our simulations. For the stop band attenuation of our designed PM filter we chose 80 dB which can be satisfied with the filter order of $N_{up}K$ = 32 that we chose, where $K$ is the length of the pulse in symbols.

In Figure 10 we show the PSD of the transmitted waveforms using SRRC and our proposed PM filter. In this Figure we also show the PSD of transmitted waveforms with higher symbol rates (16 and 20 ksps). We note that QAM and APSK modulations exhibit the same spectra, as the spectrum depends upon the filter and not the type of Q-APSK modulation.

According to the results, we notice there is significantly lower out-of-band (OOB) power emission (which directly translates to less interference on adjacent VDL2 channels) using the proposed PM filter comparing to SRRC. We note that the main lobe of the PSD increases with symbol rate, as expected. Hence, one thing that we should consider is that the total available channel bandwidth of VDL2 is 25 kHz.

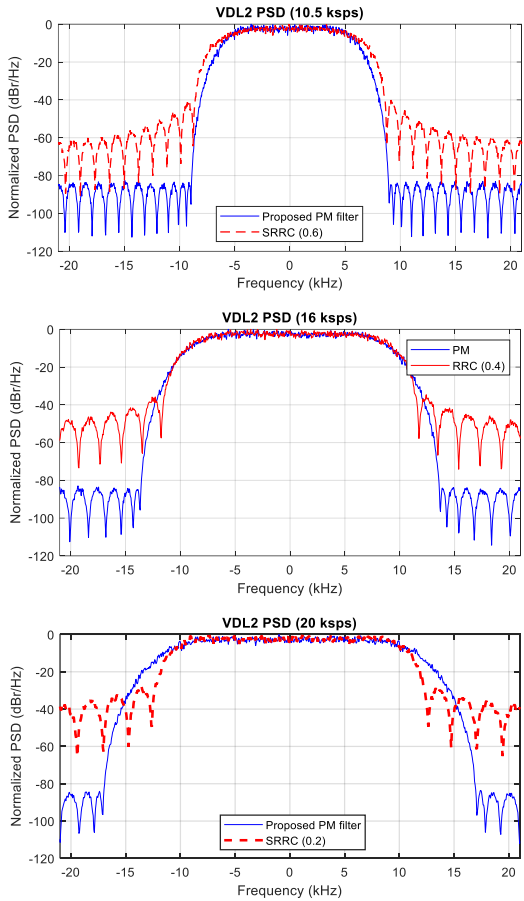

Figure 10. PSD of A-VDL with different symbol rates and pulse shaping filters.

According to these results PM filter has smaller main lobe than the VDL2 signal for symbol rate of 10.5 ksps. Therefore, this filter is outperforming SRRC for VDL2 type use. For symbol rate of 16 ksps the filter main lobe is still comparable with SRRC filter main lobe with smaller SRRC roll-off factor of $\alpha = 0.4$. The reason we decreased the $\alpha$ is to fit the main lobe of the PSD inside the 25 kHz channel, following the analysis in [1]. In Figure 10 we realize that using higher symbol rates of 20 ksps, the PM filter will not satisfy the VDL2 spectral mask requirements. Hence other filters might be designed is larger channel bandwidths cannot be allocated.

Referring to the curves shown in Figure 9, one can think of vocoders, which are another factor that play an important role on information rate and the required channel capacity for VHF voice communication. Ideally, vocoders with lower bit rates and as high a voice quality as possible are of interest. As mentioned, the vocoders used in current ACARS aviation VHF voice communications have 2400 bps bit rate. Recently more efficient algorithms and techniques were proposed that can reduce the bit rate by factor of 4 compared to ACARS vocoders with providing almost the same voice quality. Example of these vocoders are those based on CODEC 2 with minimum bit rate of 700 bps [6], Mixed Excitation Linear Predictive (MELPe) with minimum bit rate of 600 bps [7], and model - advanced multiband excitation (AMBE) developed by Digital Voice Systems, Inc. (DVSI) with minimum bit rate of 2 kbps [8]. Referring to [6]-[8] one can compare the quality of voice using these vocoders. Hence, using these lower rate vocoder techniques might be another possible solution to increase throughput within the occupied band of the future VHF voice channels.

## V. VDL Implementation Using 5G Techniques

Single carrier frequency division multiple access (SC-FDMA) using DFT-s-OFDM modulation has drawn great attention as an attractive alternative to OFDMA since its use in 4G mobile communications standard Long-Term Evolution (LTE), primarily in the uplink communications where PAPR greatly benefits the mobile terminal in terms of transmit power efficiency and reduced cost of the power amplifier. Following 4G, SC-FDMA was also adopted as one of the uplink multiple access scheme options in 5G. Here we show that using SC-FDMA we can generate the same A-VDL waveforms with comparable complexity, with approximately two times higher complexity than the conventional single carrier. In Figure 11 we show the block diagram of SC-FDMA proposed for implementing the single carrier waveforms such as A-VDL.

In Figure 11, as mentioned before, $N$ is the number of modulated symbols. For testing the SC-FDMA block diagram, we simulated the waveform with $N = 16$, $N_{up} = 4$, $K = 8$, using SRRC filter with $\alpha = 0.6$, and BSPK symbol mapping. We also generated the same symbols by convolving the samples through the filter after upsampling the symbols (as described in Figure 4). Hence, the results shown in Figure 12, confirm the identical waveform samples generated by both methods.

Here we compare the complexity of two methods. In Figure 11, the blocks which require most of computations, including $(N + K)$-Point DFT/IDFT, filtering (transmitter and receiver), and $(N + K)N_{up}$-point IDFT/DFT are shown with red color.

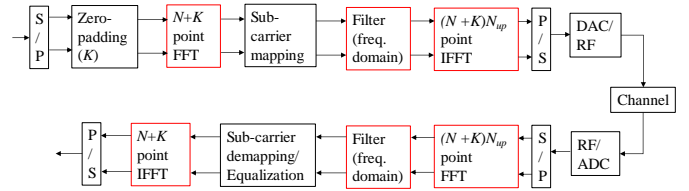

Figure 11. DFT-s-OFDM implementation of A-VDL.

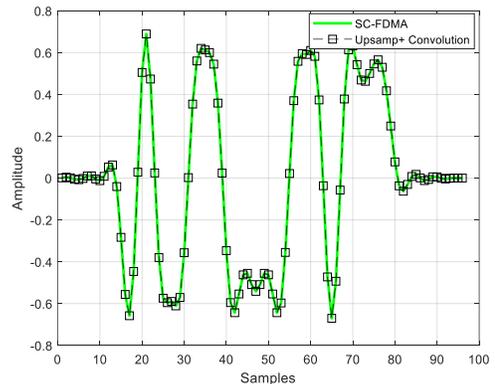

Figure 12. Identical waveforms generated by two methods

It is known that for a size of $N$ complex vector, it takes $N \log_2(N)/2$ complex multiplications to compute the fast Fourier transform (FFT) or inverse FFT (IFFT). Therefore, looking at Figure 11 one can find the overall transmitter and receiver multiplications as $[(N + K)\log_2(N + K) + (N + K)N_{up} \log_2((N + K) N_{up}) + 2(N + K) N_{up}]$, as the first part is for $(N + K)$-Point DFT/IDFT, the second part for $(N + K)N_{up}$-point IDFT/DFT, and the third part for filtering in frequency domain which requires $2(N + K) N_{up}$ number of complex multiplications for each transmitter and receiver part. In conventional single carrier such as VDL2 or our A-VDL schemes, complexity is mainly due to the filter convolution operation, which for size of symbols $N$ and filter size $L = K N_{up} + 1$, the number of complex multiplications is $[K (K +1) + K (N N_{up} - L)] / 2$. Therefore, for both transmitter and receiver we have $[K (K + 1) + K (N N_{up} - L)]$ number of complex multiplications.

As analytical results, choosing $N_{up} = 4$ and $K = 8$ we plot the number of multiplications of both schemes in Figure 13 for different number of symbols $N$. As for complexity simulations, we measured the elapsed simulation time for both methods in MATLAB, which results are shown in Figure 14. As we see, simulation results follow the same trend as analytical results, therefore confirming our complexity analysis.

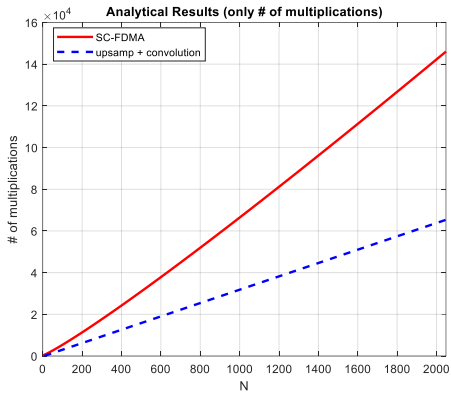

Figure 13. Number of complex multiplications of conventional filtering and SC-FDMA.

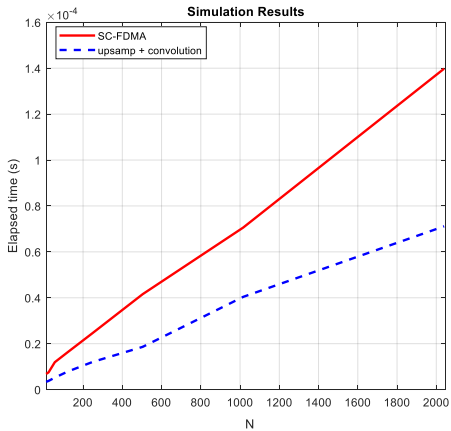

Figure 14. Elapsed time of conventional filtering and SC-FDMA (MATLAB simulations averaged over 100000 trials).

These results show that the complexity of SC-FDMA is approximately two times higher than the conventional single-carrier scheme, with modest complexity for low number of symbols (i.e., $N < 200$). We should note that SC-FDMA in comparison to conventional single carrier does have some interesting advantages such as allowing FD multiple access within the allocated channels, and if needed, simpler frequency domain channel equalizations.

## VI. CONCLUSION

In this paper we investigated several advanced technologies for aviation VHF communications, or VDL, which we term, as in [1], Advanced-VDL (AVDL). Proposing an updated VDL frame structure, we show how novel techniques such as LDPC channel encoding can significantly improve both power and spectral efficiency. In addition, we investigated more efficient standardized voice encoders, as well as better pulse shaping filters than the classical SRRC filter used in VDL. Via simulations and analysis, we compared the proposed schemes' PAPR, spectral efficiency, and link margin compared to VDL2/3 for different QAM and APSK modulation orders; we also included coherent 8-PSK. Our results show that using modulation orders of 256-QAM or 256-APSK we can increase the useful information bit rate by at least a factor of 2, with a very slight reduction on link margin, 0.8 dB for 256-QAM and 1.8 dB for 256-APSK. We proposed a filter than can decrease the out-of-band power emission of A-VDL significantly, which can reduce the A-VDL adjacent channel interference, further improving system spectral efficiency. In addition, as another way of implementing A-VDL we proposed and investigated the single-carrier type waveform SC-FDMA used in cellular LTE and 5G uplink communication links. According to our analysis and results, we achieve comparable complexity, with approximately two times higher complexity than conventional single-carrier A-VDL. Therefore, implementing A-VDL based on SC-FDMA might be of interest for future single carrier aviation communications, considering the multiple access advantage of SC-FDMA. As future work, we can do some lab testing after implementing A-VDL on software define radios (SDRs), and actual flight testing.